\documentclass[preprint,3p,11pt,onecolumn]{elsarticle}
\usepackage{natbib} 
\usepackage{amssymb}
\usepackage{amsmath} 
\usepackage{caption}
\usepackage{graphicx}
\graphicspath{{./figures/}}
\usepackage{xcolor}

\usepackage[colorlinks,allcolors=blue]{hyperref}

\usepackage{subcaption}
\usepackage{adjustbox}
\usepackage{svg}

\definecolor{bleudefrance}{rgb}{0.19, 0.55, 0.91}

\usepackage{color}

\journal{Powder Technology}

\usepackage{siunitx}

\begin{document}

\begin{frontmatter}



\title{Shape effects in binary mixtures of PA12 powder in additive manufacturing}
\author{Sudeshna Roy}
\author{Thorsten P\"oschel}
\address{
Institute for Multiscale Simulation, Friedrich-Alexander-Universit\"at Erlangen-N\"urnberg, Erlangen, Germany}
\cortext[correspondingauthor]{Correspondence: sudeshna.roy@fau.de (S.R.)}

\begin{abstract}
The quality of the powder spread in additive manufacturing devices depends sensitively on the particles' shapes. Here, we study powder spreading for mixtures of spherical and irregularly shaped particles in Polyamide 12 powders. Using DEM simulations, including heat transfer, we find that spherical particles exhibit better flowability. Thus, the particles are deposited far ahead of the spreading blade. In contrast, a large fraction of non-spherical particles hinders the flow. Therefore, the cold particles are deposited near the front of the spreading blade. This results in a temperature drop of the deposited particles near the substrate, which cannot be seen with spherical particles. The particles of both shapes are homogeneously distributed in the deposited powder layer.
\end{abstract}

\end{frontmatter}


\section{Introduction}
Polyamide 12 (PA12) is a widely applied material for additive manufacturing due to its exceptional durability, chemical resistance, flexibility, low friction, ease of processing, wide availability, and, in specific formulations, biocompatibility \cite{tan2020recent,priyadarshini20233d}.
Its toughness makes it ideal for producing durable parts, while its chemical resistance is advantageous in industries such as healthcare and manufacturing \cite{zhu2017thermal}. 
Additionally, its flexibility is suitable for applications that require parts that can bend without breaking, and its low friction is valuable in reducing resistance.  PA12's ease of processing through various 3D printing techniques and its accessibility in different forms contribute to its popularity. Therefore, PA12 is widely used as a material of choice for applications where a combination of strength, chemical resistance, and flexibility is crucial. 

Fresh PA12 powder exhibits a wide range of size and morphology, with particles having rough surfaces and random ellipsoidal shapes \cite{simha2021polyamide}. These diverse irregular shapes in PA12 powder have a significant impact on the density and structure of the packing. Incorporating non-spherical particles into a polymer powder matrix offers distinct benefits regarding heat transfer. The irregular geometry of these particles leads to an increased number of contact points, facilitating enhanced thermal conductivity and heat transfer pathways compared to their spherical counterparts \cite{gan2017effect,govender2020effect}. Furthermore, the irregular arrangement of non-spherical particles results in improved packing density \cite{nasato2020influence}, thereby enhancing overall heat conduction within the powder bed. Additionally, the anisotropic heat transfer characteristics introduced by non-spherical particles can be used to direct the heat flow as per specific requirements. Although irregularly shaped particles offer a larger exposed surface area, leading to improved heat conductivity within the powder bed, they come with inherent disadvantages, notably reduced flowability. This limitation challenges the efficiency of additive manufacturing processes, particularly in applications such as laser beam melting, where factors such as processing speed and material selection are crucial \cite{goodridge2012laser}. To improve flowability, the employment of spherical particles with a narrow size distribution is advantageous \cite{fanselow2016production,schmidt2014novel,schmidt2016optimized,blumel2015increasing,ziegelmeier2015experimental,dadbakhsh2016effect,sachs2015rounding,yang2019preparation}. These spherical particles, often obtained through innovative production methods such as melt emulsification \cite{fanselow2016production} and wet grinding with a rounding process \cite{schmidt2014novel}, are preferred in additive manufacturing processes as they improve flowability and mitigate the drawbacks associated with irregular shapes.
Thus, combining distinct shapes of PA12 particles improves the mechanical properties of printed objects, such as strength, flexibility, and durability, by creating composite materials with customized characteristics \cite{lanzl2019selective,omar2022effect}. Furthermore, this amalgamation of various particle shapes contributes to improved thermal properties \cite{lanzl2019selective}. Therefore, blends of PA12 particles of different shapes offer clear advantages in the powder-spread process, allowing the production of superior product parts. Although the effects of the morphological properties of the powder on the quality of the powder layer have been extensively studied \cite{parteli2016particle,nasato2020influence,nasato2021influence}, little is known about the relation between the shape of the particles and the thermal properties in the powder spread. Therefore, the subject of this paper is the heat transfer in the powder bed in the process of powder spreading in additive manufacturing. In particular, we study the influence of the composition of the powder from PA12 particles of different shapes.

Segregation poses a significant challenge in handling and manufacturing particulate mixtures, with notable cost implications. For example, it can lead to increased expenses due to the loss of homogeneity of the mixture, which can result in customer dissatisfaction or batch failure during quality assurance checks. Internal factors such as variations in size, density, shape, and surface properties are primary contributors to segregation. Shape-induced segregation and inhomogeneity are frequently observed in granular mixtures across industrial and geophysical applications \cite{alizadeh2017effect, shimosaka2013effect, he2021particle, cunez2024particle}. Discrete Element Method (DEM) simulations are commonly used to simulate complex shapes of particulate systems in these studies efficiently. Therefore, we employ DEM simulations to investigate the effects of particle shape in mixtures on powder flowability, local structural homogeneity, and the temperature profile of the powder layer in additive manufacturing. 

In this paper, we study the influence of the granular mixture's composition of irregularly shaped particles and spheres with regard to the properties of the powder spread in additive manufacturing devices, and the quality of the resulting deposited powder layer. We model three-dimensional particles of realistic irregular shapes using the multisphere approach \cite{Buchholtz:1993}. For the simulation of the powder spreading, we employed such mixtures of identical volume in varying fractions. Newton's equations were integrated using open source code \textsc{MercuryDPM} enhanced by a thermal conductivity module \citep{roy2024combined}. We focus on three details important to powder-based additive manufacturing processes: (1) the particle velocity profile in front of the powder spreading blade; (2) the particle temperature beneath the moving blade during the process, and (3) the distribution of particles of different shape in the deposited powder layer. 

\section{Methods}
\subsection{Multisphere modeling of PA12 particles}

\begin{figure}
    \centering
\includegraphics[width=0.5\columnwidth,bb=10 150 365 290,clip]{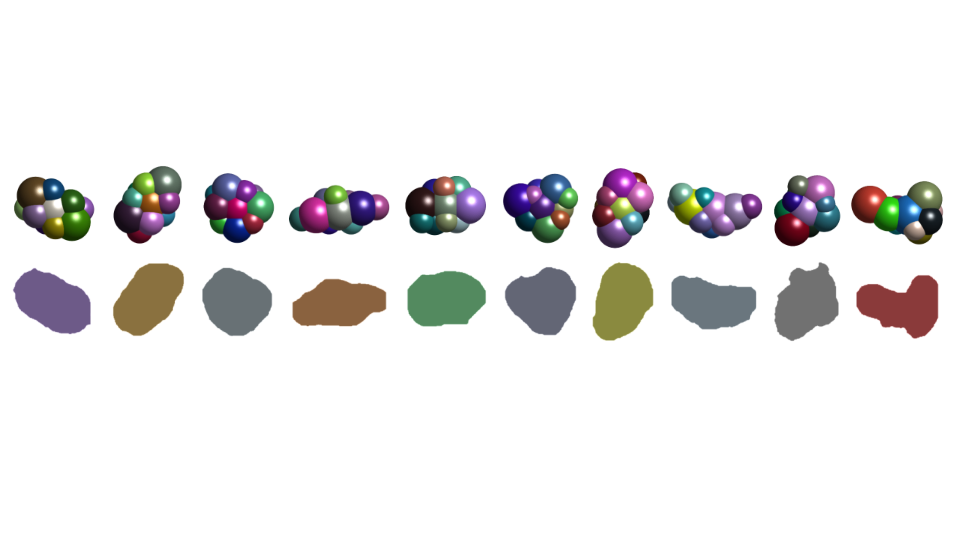}
\includegraphics[width=0.5\columnwidth,bb=375 150 710 290,clip]{PA12_particles.png}
    \caption{Examples of non-spherical multisphere particles used in the DEM simulation. Each multisphere consists of 10 spheres arranged to approximate as close as possible the shape of microscopic SEM images of PA12 powder particles, also shown.}
    \label{fig:PA12_particles}
\end{figure}

For realistic modeling of the particle's shapes, we start from SEM images of PA12 particles and transform them into three-dimensional representations \citep{nasato2021influence}. These 3D objects serve us as the target geometries for generating multisphere particles by means of the generator \textsc{Clump} \cite{angelidakis2021clump}. Each multisphere particle consists of 10 spheres to approximate the non-spherical target particle shapes. \autoref{fig:PA12_particles} shows examples of such particles. 
The volume of the non-spherical particles is the same as the volume of the spherical particles we use for comparison. We study binary mixtures of non-spherical and spherical PA12 particles subjected to the powder spreading process, with the fraction of non-spherical particles $\alpha\in[0,1]$. Thus, $\alpha$ characterizes the composition of the powder with respect to shape. Therefore, we study the properties of the spreading process and of the resulting powder layer in dependence on $\alpha$.

\subsection{Thermo-mechanical contact model}
The total normal force between spheres in contact, $\vec{F}_n$, is given by the sum of the viscoelastic force $\vec{F}^{\mathrm{el}}_n$ and the adhesive force $\vec{F}^{\mathrm{adh}}_n$. The viscoelastic contact is described by the dissipative Hertz contact law \cite{BSHP:1996,Algo:2005}, using temperature-dependent material properties, 
\begin{equation}
    \vec{F}^{\mathrm{el}}_n = \min\left(0, -\rho \xi^{3/2} - \frac{3}{2}A_n\rho\sqrt{\xi}\dot{\xi}\right) \vec{e}_n\,,
    \label{Eq:Hertz_normal}
\end{equation}
where $\vec{e}_n$ is the unit vector between the centers of the involved spheres. The elastic constant is given by \cite{BSHP:1996}
\begin{equation}
    \rho \equiv \frac{4}{3} \, E^{*}_{T} \, \sqrt{R^*} \,,
\end{equation}
with the temperature dependent effective Young's modulus $E^*_T$ and the effective radius $R^*\equiv R_1R_2/\left(R_1+R_2\right)$. 
For identical materials of the particles in contact, the elastic constant simplifies to $\rho = \frac{2E\sqrt{R^*}}{3\left(1-\nu^2\right)}$, where $E$ and $\nu$ are the Young's modulus and Poisson ratio of the material, respectively. Since \autoref{Eq:Hertz_normal} corresponds to an impact velocity dependent coefficient of restitution \cite{Schwager:1998, Schwager:2008}, in agreement with experimental evidence \cite{Bridges:1984, Hatzes:1988}, we determine the dissipative constant, $A_n$, using the method described in Ref. \cite{muller2011collision}, where we assume the coefficient of restitution $0.7$ for the impact velocity $50$ mm/s. The full list of model parameters is provided in \autoref{tab:modelling_parameters}.
\begin{table}[ht!]
    \centering
    \caption{Material and model parameters together with the corresponding references}
    \label{tab:modelling_parameters}
    \begin{tabular}{||c | c | c||} 
    \hline
    constant & value & reference \\ [0.5ex] 
    \hline\hline
    particle diameter $D_{50}$ & $29.6 ~\mu$m & \cite{parteli2016particle,nasato2021influence} \\ 
    density $\rho_m$ & $1000$ kg/m$^{-3}$ &  \cite{nasato2021influence} \\
    reduced Young's modulus $E_i, E_j$ & $2.3\times{10}^6$ Pa &  \cite{salmoria2012mechanical,parteli2016particle} \\
    physical Young's modulus $E_\mathrm{ph}$ & $2.3\times{10}^9$ Pa &  \cite{salmoria2012mechanical,parteli2016particle} \\
    Poisson ratio $\nu_i, \nu_j$ & $0.40$ &  \cite{nasato2021influence} \\
    dissipative constant $A_n$ & $1.85\times{10}^{-7}\text{s}$ & \cite{muller2011collision} \\
    Coulomb's friction coefficient $\mu$ & $0.50$ & \cite{nasato2021influence} \\
    latent heat of fusion $L_f$ & $101.66\times{10}^3$ J/kg& \cite{kulinowski2022development} \\ 
    solid heat capacity $c_p$ & $1200$ J/{kgK} & \cite{hejmady2019novel} \\ 
    thermal conductivity coefficient $k_\mathrm{ph}$ & $0.12$ W/mK & \cite{hejmady2019novel,yuan2011thermal,laumer2014fundamental} \\
    thermal convection coefficient $h_\mathrm{conv}$ & $15$ W/m$^2$K & \cite{riedlbauer2015thermomechanical,peyre2015experimental,yaagoubi2021review} \\ 
    emissivity $\epsilon$ & $0.80$ & \cite{,yaagoubi2021review,belliveau2020mid,peyre2015experimental} \\ 
    melting temperature $T_m$ & $451.15$ K & \cite{hejmady2019novel} \\ 
    temperature interval $\Delta{T}$ & $20$ K & \cite{ganeriwala2016coupled} \\ 
    substrate temperature $T_0$ & $443$ K & \cite{peyre2015experimental,laumer2015influence} \\
        initial particle temperature $T_\mathrm{in}$ & $393$ K & \cite{peyre2015experimental,li2021experimental} \\ [1ex] 
    \hline
    \end{tabular}
\end{table}

The dependence of the particle material on the temperature is modeled through temperature-dependent Young's modulus,
\begin{equation}
    E^{*}_{T} =\frac{E^{*}_0}{2} \left[1+\tanh{\left(\frac{T_m-T}{\Delta{}T}\right)}\right]\,,
\label{eq:functionModulus}    
\end{equation}
where $T_m$ is the melting temperature, $T\equiv\frac{T_i+T_j}{2}$ is the average temperature of the particles in contact, and $\Delta{}T$ is the width of the transition zone (see \cite{luding2005discrete, shaheen2024numerical} for details).  
The effective Young's modulus at room temperature is \cite{Algo:2005}
\begin{equation}
    E^{*}_{0}=\left(\frac{1-\nu_i^2}{E_i} + \frac{1-\nu_j^2}{E_j}\right)^{-1}\,.
\end{equation}
where the indices $i$ and $j$ refer to the involved spheres of different materials. 

Taking into account the temperature dependence of the material characteristics is important for additive manufacturing processes, as the temperature changes considerably during the coating process and can be close to  $T_m$. 

The tangential force between spheres in contact,
\begin{equation}
    \vec{F}_t = -\min \left[ \mu\left|\vec{F}_n\right|,  \int_\text{path}^{} 8\, G^{*}_{T}\sqrt{R^* \xi} \,\text{d}s 
    + A_t  \sqrt{R^* \xi}\, v_t \right] \vec{e}_t
    \label{eq:Ft}
\end{equation}
extends the model given in \citep{shaheen2024numerical} to a fully-temperature-dependent Hertz-Mindlin contact law. \autoref{eq:Ft} combines the no-slip Mindlin model \cite{mindlin1949compliance} for the elastic tangential contact force with the tangential dissipative force given in \cite{parteli2016particle} with $A_t \approx 2 A_n E^{*}_{T}$. The temperature-dependent shear modulus is \cite{roy2024combined}
\begin{equation}
    G^{*}_{T} =\frac14 \kappa E^{*}_{T}\,,
\end{equation}
where $\kappa$ is the stiffness ratio, defined for two identical materials with Poisson's ratio $\nu$, given as:
\begin{equation}
\kappa\equiv\frac{2\left(1-\nu\right)}{2-\nu}\,.
\end{equation}

The adhesive force between spheres in contact is described by the Johnson-Kendall-Roberts (JKR) model \citep{johnson1971surface,JKR-BASP:2007,JKR-BASP:2007E},
\begin{equation}
    \vec{F}^\text{adh}_n = 4\sqrt{\pi a^3\gamma E^{*}_{T}} \; \vec{e}_n\,,
\end{equation}
where $\gamma$ is the surface energy density and $a$ is the contact radius due to the mutual particle deformation. It relates to the deformation $\xi$ via  
\begin{equation}
    \xi = \frac{a^2}{R^*} - \sqrt{\frac{4\pi a\gamma}{E^*}}\,.
\end{equation}
The maximum interaction distance at which the contact breaks under tension is  
\begin{equation}
\xi_t = \frac{1}{2}\frac{1}{6^{1/3}}\frac{a^2}{R^*}\,.    
\end{equation} 
The surface energy density of PA12 is $32$\,mJ/m$^2$ \cite{bahrami2022experimental,novak2006surface}. Following Nasato et al.~\cite{nasato2021influence}, for the simulation we reduce $\gamma$ by the same factor as the Young's modulus to achieve a realistic angle of repose. We checked that $\gamma=0.01$\,mJ/m$^2$  gives the angle of repose $\approx 30^{\circ}\dots  35^{\circ}$, which is a realistic value for PA12 at room temperature.

 \subsection{Heat transfer for multisphere particles}

\begin{figure}[hbt!]
\centering
{\includegraphics[trim={0cm 0.5cm 0cm 0.7cm},clip,width=0.5\columnwidth]{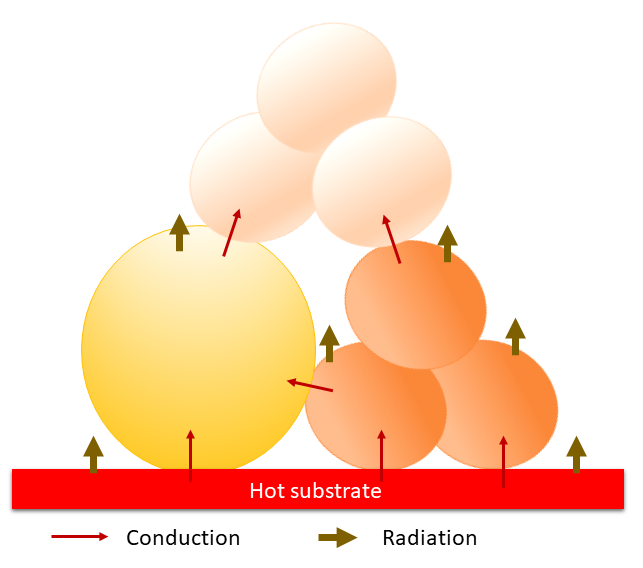}}
\caption{Sketch of the heat transfer mechanisms \textit{conduction} and \textit{radiation} in mixtures of spherical and non-spherical particles. The same color indicates spheres that belong to the same spherical/non-spherical particle.}
\label{fig:HeatTransfer}
\end{figure}

Heat transfer occurs via three primary mechanisms, conduction, convection, and radiation. Conduction is active when a sphere of a multisphere particle contacts a sphere of another particle or the boundary. Convection and radiation occur fat the surface of a particle, facilitating heat transfer to surrounding particles and the ambient. Here, we consider the spread of pre-heated powder over the surface of the printed part at melting temperature. In this case, the temperature of the powder and the ambient temperature are assumed identical, and major heat exchange occurs between the particles and the solidified part. We do not consider thermal convection here.
Each particle, $i$, is assigned a uniform temperature, $T_i$, which varies over time according to
\begin{equation}
     \frac{\partial T_i}{\partial t}=\frac{Q_i}{m_i\, c_p},
\label{eq:temp}
\end{equation}
where $c_p$ is the heat capacity. The total thermal transfer rate reads
\begin{equation}
     Q_i=\sum_{j=1}^{N}Q^{\mathrm{cond}}_{ij} + \sum_{j=1}^{N}Q^{\mathrm{rad}}_{i,j} + Q^{\mathrm{rad}}_{i,0}
\label{eq:therml}
\end{equation}
where $Q^{\mathrm{cond}}_{ij}$ and $Q^{\mathrm{rad}}_{i,j}$ are the rate of heat conduction and the rate of radiation heat exchange between particles $i$ and $j$, and $Q^{\mathrm{rad}}_{i,0}$ is the rate of radiation heat exchange between particle $i$ and the bottom substrate. A sketch of the heat exchange system is shown in \autoref{fig:HeatTransfer}. 

The conduction rate between multisphere particle $i$ and $j$ results from the contributions of any of the $N^{(i)}$ spheres that constitute $i$ in contact with any of the $N^{(j)}$ spheres that constitute $j$:
\begin{equation}
     Q^{\mathrm{cond}}_{ij}= \frac{T_j-T_i}{\Delta_{ij}} \sum_{u=1}^{N^{(i)}}\sum_{v=1}^{N^{(j)}} k\, s_{uv} \,,
\label{eq:thermlA}
\end{equation}
where $k$ is the thermal conductivity, and $s_{uv}$ is the contact area of spheres $u$ (being part of particle $i$) and sphere $v$ (belonging to particle $j$). Note that only spheres $u$ and $v$ in contact contribute to the sum since only for such particles $s_{uv}\ne 0$.

The temperature gradient in \autoref{eq:thermlA} is calculated on the multisphere particle level, where $T_i$ is the temperature of multisphere particle $i$ and $\Delta_{ij}$ is the distance between the centroids of the particles $i$ and $j$. 

In DEM simulations, we have to reduce the modulus of elasticity in order to achieve a realistic calculation time. This leads to an unrealistically large contact area between the spheres, which distorts the heat conduction rate between particles in contact. For this reason, we must not use the physical heat conduction coefficient $k_\text{ph}$ given in \autoref{tab:modelling_parameters} directly.
Instead, we use the corrected value
\begin{equation}
    k=k_\text{ph} \left(\frac{E}{E_\text{ph}}\right)^\frac{2}{3}\,.
\end{equation}
The exponent 2/3 exponent originates from the relation between the contact area and the Young's modulus.

For the heat transfer rate due to radiation we follow similar arguments as for $Q^{\mathrm{cond}}_{ij}$: We assume that each sphere $u$ belonging to particle $i$ exchanges heat through radiation with each sphere $v$ belonging to particle $j$, provided the distance of $u$ and $v$ is less than $1.5D_{50}$ \cite{zhou2023cfd,zhu2020130}. The heat transfer between the spheres is then governed by the Stefan-Boltzmann law:
\begin{equation}
     Q^{\mathrm{rad}}_{ij}= \sigma_\text{SB} \, \epsilon \, \left(T_i^4-T_{j}^4\right)\, \sum_{u=1}^{N^{(i)}}\sum_{v=1}^{N^{(j)}} A_{uv} \,,
\label{eq:therml-rad}
\end{equation}
%
%
with the Stefan-Boltzmann constant $\sigma_\text{SB}=5.67\times 10^{-8}\,\text{W}/\left( \text{m}^2/\text{K}^4\right)$, the material emissivity $\epsilon$, and the effective area of radiation between two the spheres $u$ and $v$ 
\begin{equation}
    A_{uv}=
\begin{cases}
\frac{1}{2}\left(\frac{A_i{r_u}^2}{{R_i}^2}+\frac{A_j{r_v}^2}{{R_j}^2}\right)G(\left|\vec{r}_u-\vec{r}_v\right|) & \text{if}~\left|\vec{r}_u-\vec{r}_v\right| < 1.5D_{50}\\
0 & \text{otherwise}\,.
\end{cases}
\end{equation}
where $R_i$ and $R_j$ are the mean radius of the multisphere particles $i$ and $j$, respectively, $r_u$ and $r_v$ are the radius of the spheres $u$ and $v$, $A_i$ and $A_j$ are the surface areas of particles $i$ and $j$, respectively, and $G(\left|\vec{r}_u-\vec{r}_v\right|)$ is a shielding factor accounting for blockage of radiation between particles. The shielding factor is given as 
\begin{equation}
    G(\left|\vec{r}_u-\vec{r}_v\right|) = \frac12 \left[1-\text{erf}\left(\frac{\left|\vec{r}_u-\vec{r}_v\right|-D_{50}}{D_{50}/16}\right)\right]\,,
\end{equation}
where $D_{50} = 29.6 \,\mu$m is the mean diameter of the non-spherical particles and $D_{50}/16$ is the width of the error function which is considered here based
on the requirement of a smooth transition of the radiation heat transfer across the bed in the $z$ direction. 
For the radiation heat exchange between particle $i$ and the hot substrate, we write 
\begin{equation}
Q^{\mathrm{rad}}_{i,0} = \sigma_{SB}\,\epsilon\, \left(T_i^4-T_0^4\right)\,\sum_{u=1}^{N^{(i)}} A_{u} \,,
\end{equation}
where $T_0$ is the temperature of the substrate and the effective area of radiation $A_u$ is approximated by 
\begin{equation}
    A_{u}=
\begin{cases}
\left(\frac{A_i{r_u}^2}{{R_i}^2}\right)G(z_u) & \text{if}~z_u < 1.5D_{50}\\
0 & \text{otherwise}\,.
\end{cases}
\end{equation}
where $z_u$ is the distance between particle $u$ and the substrate and $G(z_u)$ is a 
shielding factor for radiation between particle $i$ and the substrate given as


\begin{equation}
    G(z_u) = \frac12 \left[1-\text{erf}\left(\frac{z_u-D_{50}}{D_{50}/16}\right)\right]\,,
\end{equation}


\subsection{Numerical setup}

\begin{figure}[htb!]
\centering
\begin{subfigure}{\columnwidth}
    \includegraphics[trim={0cm 0cm 0cm 0cm},clip,width=\columnwidth]{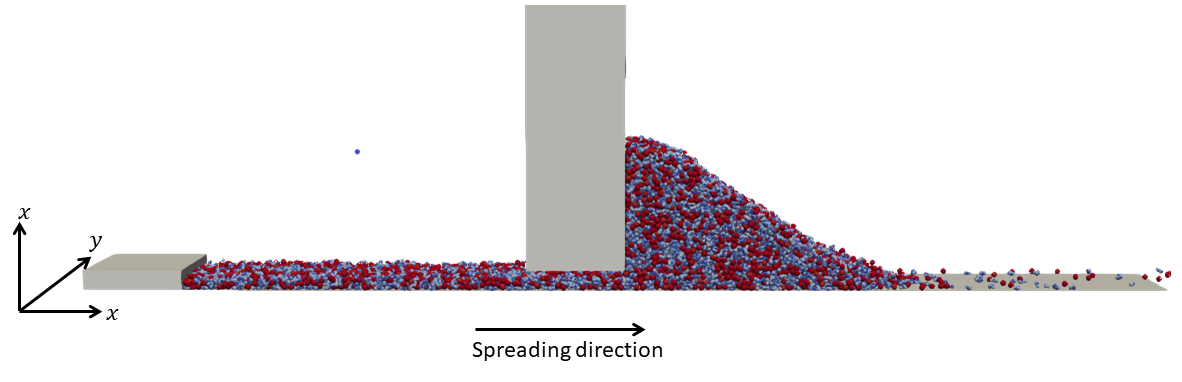}
    \subcaption{}
\end{subfigure}
\hfill
\begin{subfigure}{0.48\columnwidth}
    \includegraphics[trim={0cm 0cm 0cm 0cm},clip,width=\columnwidth]{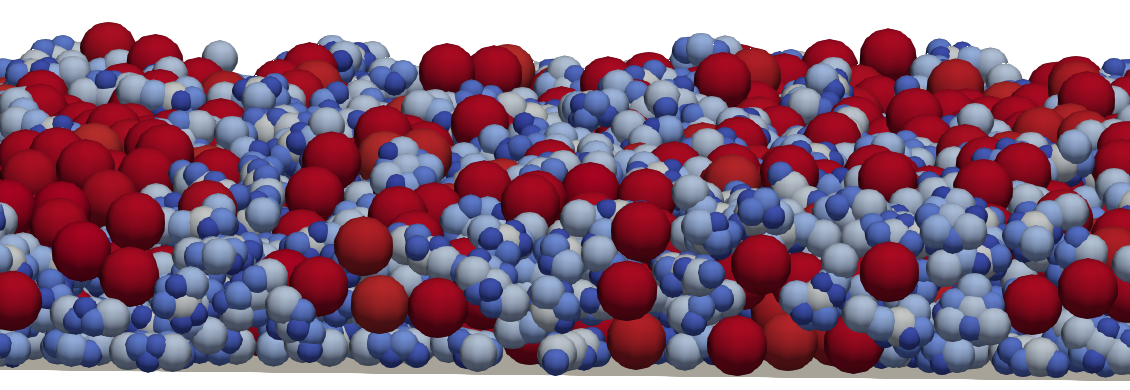}
  \hfill  
      \includegraphics[trim={0cm 0cm 0cm 0cm},clip,width=\columnwidth]{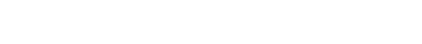}    
    \subcaption{}
\end{subfigure}
\hfill
\begin{subfigure}{0.48\columnwidth}
    \includegraphics[trim={0cm 0cm 0cm 0cm},clip,width=\columnwidth]{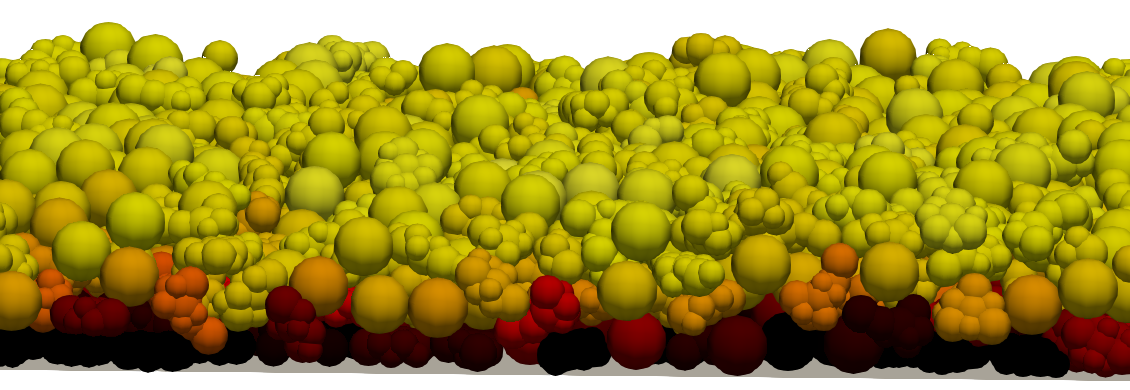}
  \hfill  
      \includegraphics[trim={0cm 0cm 0cm 0cm},clip,width=\columnwidth]{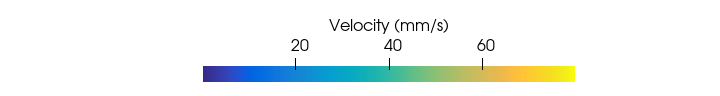}
    \subcaption{}
\end{subfigure}
\caption{Numerical setup: (a) A powder mixture of volume fraction $\alpha = 0.6$ of spherical particles is spread. (b) deposited powder layer. Non-spherical particles are shown in blue, spherical particles are shown in red. (c) same as (b) but the color codes the temperature after $t = 0.05\,\text{s}$ of the powder spreading process. }
\label{fig:powderSpreading}
\end{figure}

The numerical setup is sketched in \autoref{fig:powderSpreading}.
A powder bed of width $0.75$ mm is simulated using periodic boundary conditions in $y$-direction. 
We insert both spherical and non-spherical particles in front of the blade, at $(x,y,z) \in [0.5,10]$ mm $\times [0,0.75]$ mm $\times [0,h]$ mm until the total bulk particle volume equals $0.70$ mm$^3$, which is sufficient material to create a powder layer of $10$ mm length, $0.75$ mm width and $0.1$ mm height. The non-spherical particles are of ten distinct shapes, as shown in \autoref{fig:PA12_particles}, and the corresponding spherical particles of equal volume have diameters within the narrow range of $[29.2,30]\,\mu\text{m}$. The total number of particles is approximately $40,000$ at an initial temperature $T_mathrm{in} = 393$ K. After the particles settled down under gravity and the system is relaxed, the spreading process starts by moving the blade at a constant velocity $50$ mm/s. \autoref{fig:powderSpreading}(a) shows a snapshot of the process. Due to this process, the particles are deposited in a layer between the blade and the substrate, where the gap height between the blade and the substrate is $H = 100 \, \mu$m, which is approximately $3.3$ times the mean diameter of the multisphere particles. The deposited powder layers are shown in \autoref{fig:powderSpreading}(b) and (c). Particles reaching the end of the powder bed (at $x = 10$ mm) are not used for the analysis. The powder layer, spread on the hot substrate, undergoes heating through both conduction and radiation. Finally, the simulation is stopped when the temperature of the powder layer reaches a steady state. The total process and, thus, the  simulation time, is $t_\mathrm{max} = 1\,\text{s}$, taking into account the time for initial settling and relaxation of particles, the spreading time, and the time for the temperature of the powder layer to achieve the steady state.

\section{Results}
\subsection{Velocity profile of the moving front}

\begin{figure}[htb!]
\centering
\begin{subfigure}{0.49\columnwidth}
    \includegraphics[trim={0cm 0cm 0cm 0cm},clip,width=\linewidth]{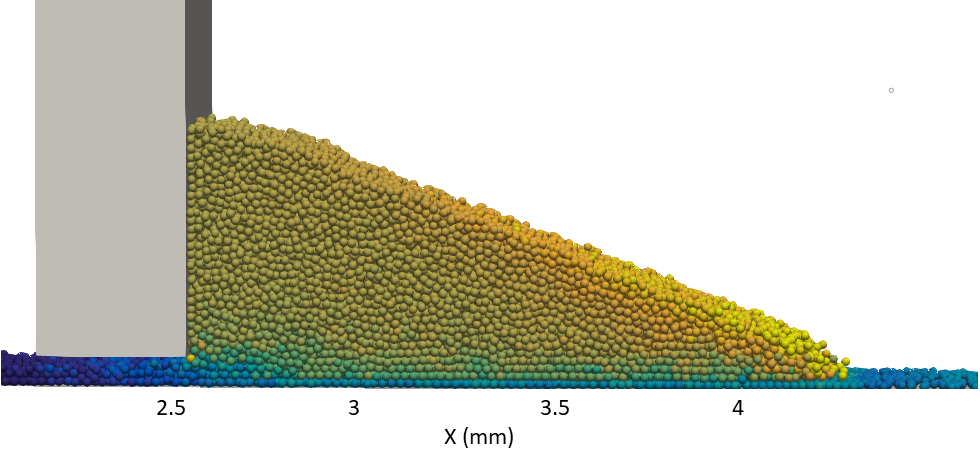}
    \subcaption{$\alpha = 0$ at $t = 0.06\text{s}$}
\end{subfigure}
\hfill
\begin{subfigure}{0.48\linewidth}
    \includegraphics[trim={0cm 0cm 0cm 0cm},clip,width=\linewidth]{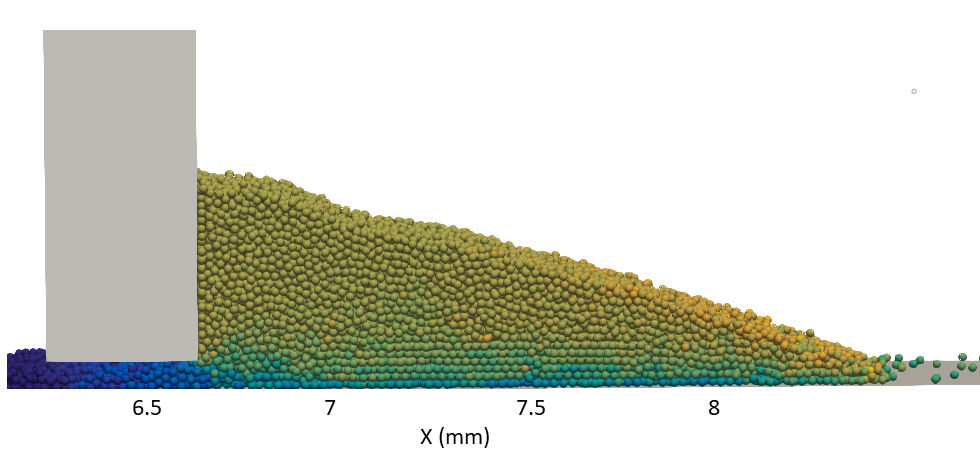}
    \subcaption{$\alpha = 0$ at $t = 0.1\text{s}$}
\end{subfigure}
\hfill
\begin{subfigure}{0.49\linewidth}
    \includegraphics[trim={0cm 0cm 0cm 0cm},clip,width=\linewidth]{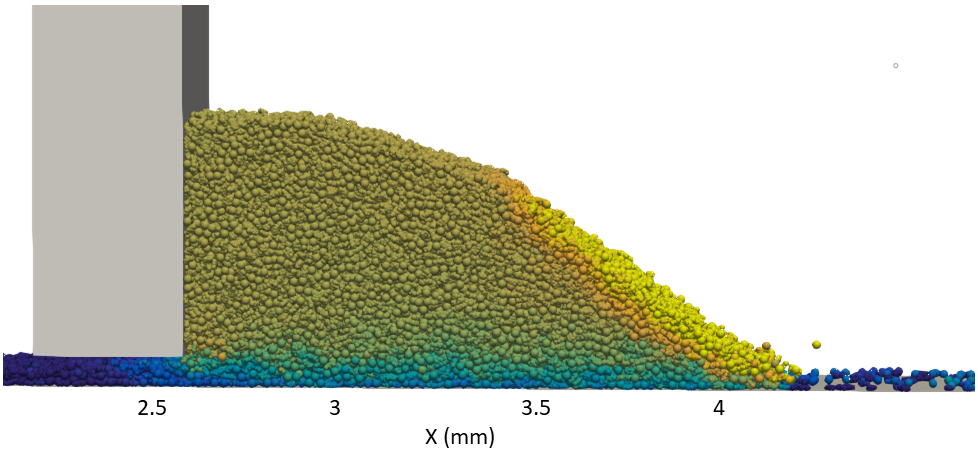}
    \subcaption{$\alpha = 0.6$ at $t = 0.06\text{s}$}
\end{subfigure}
\hfill
\begin{subfigure}{0.48\linewidth}
    \includegraphics[trim={0cm 0cm 0cm 0cm},clip,width=\linewidth]{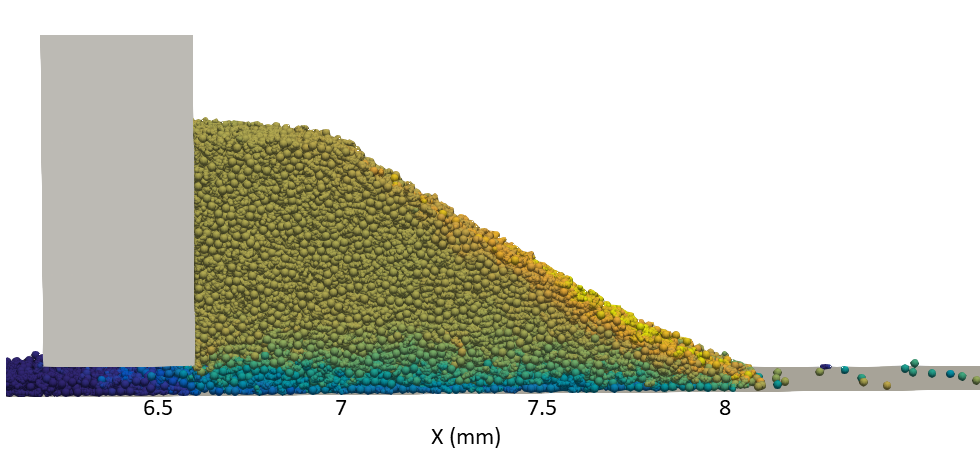}
    \subcaption{$\alpha = 0.6$ at $t = 0.1\text{s}$}
\end{subfigure}
\hfill
\begin{subfigure}{0.49\linewidth}
    \includegraphics[trim={0cm 0cm 0cm 0cm},clip,width=\linewidth]{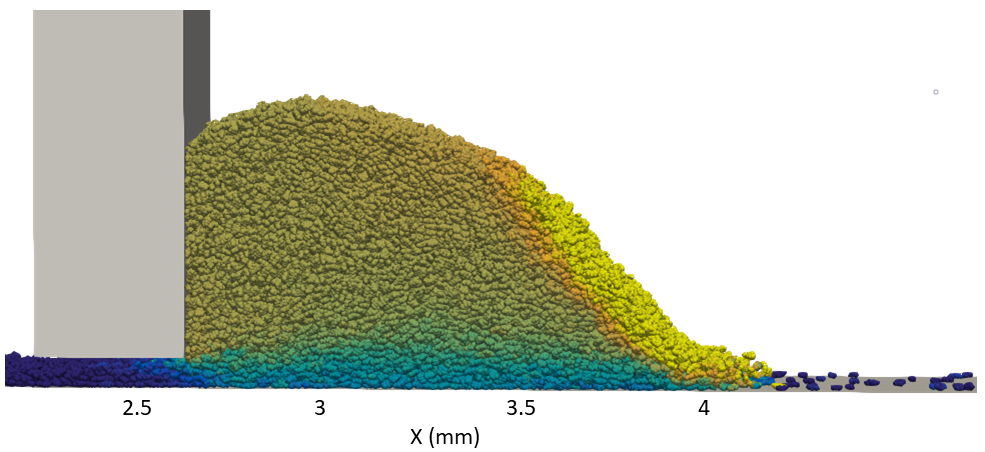}
    \subcaption{$\alpha = 1$ at $t = 0.06\text{s}$}    
\end{subfigure}
\hfill
\begin{subfigure}{0.48\linewidth}
    \includegraphics[trim={0cm 0cm 0cm 0cm},clip,width=\linewidth]{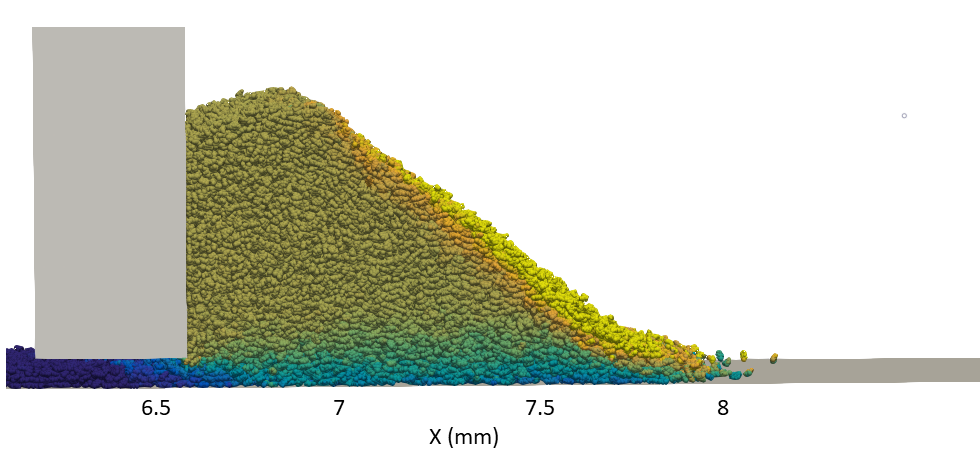}
    \subcaption{$\alpha = 1$ at $t = 0.1\text{s}$}
\end{subfigure}
\hfill
\begin{subfigure}{\linewidth}
    \includegraphics[trim={0cm 0cm 0cm 0cm},clip,width=0.8\linewidth]{figures/Colorbar.png}
\end{subfigure}
\caption{Snapshots of the powder spreading process for binary mixtures of volume fraction $\alpha$ at time $t$. The color of the particles codes their velocity}
\label{fig:VelocityX}
\end{figure}

\autoref{fig:VelocityX}(a-f) 
shows snapshots taken at time $t = 0.06\,\text{s}$ and $t = 0.1\,\text{s}$ of the powder spreading process for mixtures with volume fraction $\alpha \in\{0, 0.6, 1\}$. The color of the particles indicates their current velocity. For all mixtures, the particles on the free surface ahead of the moving blade move with larger velocities at $t = 0.06\,\text{s}$ (\autoref{fig:VelocityX}(a,c,e)) compared to $t = 0.1\,\text{s}$ (\autoref{fig:VelocityX}(b,d,f)). Increasing the fraction of nonspherical particles (increasing $\alpha$) reduces the flowability of the powder. Therefore, with increasing $\alpha$, the slope of the heap in front of the moving blade increases as well.
\begin{figure}[htb!]
\centering
\begin{subfigure}{0.48\linewidth}
    \includegraphics[trim={0cm 0cm 0cm 0cm},clip,width=\linewidth]{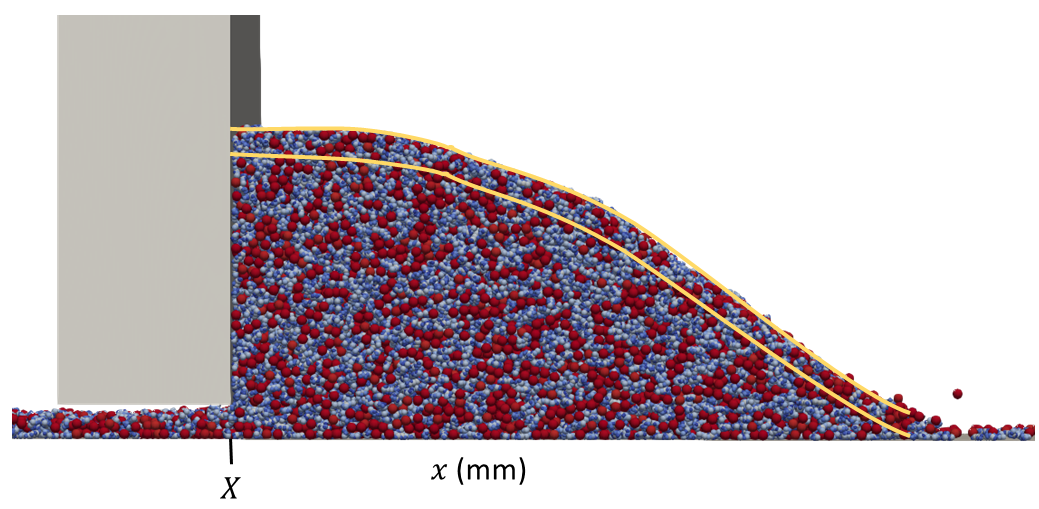}
    \subcaption{}
\end{subfigure}
\hfill
\begin{subfigure}{0.48\linewidth}
    \includegraphics[trim={0cm 0cm 0cm 0cm},clip,width=\linewidth]{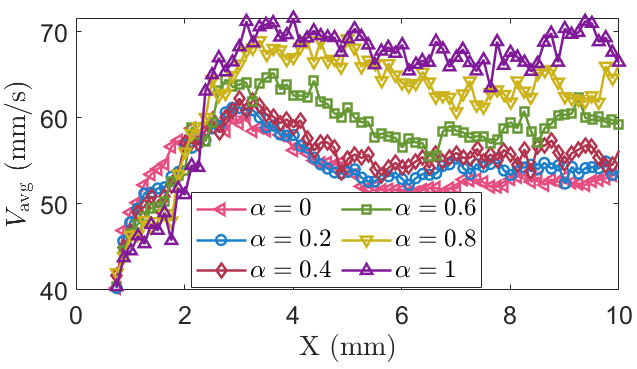}
        \subcaption{}
\end{subfigure}
\caption{(a) The yellow lines in the snapshot define the region of the surface flow. The sketch also defines the spreading length $X$. (b) Average velocity of particles near the free surface as a function of the spreading length $X$ for mixtures with different ratio $\alpha$ of spherical and non-spherical particles. }
\label{fig:VelocityProfile} 
\end{figure}

Consider the average velocity $V_\mathrm{avg}$ of the particles near the free surface for different $\alpha$. To do this, we look at the particles that are in a layer whose thickness is $4D_{50}$ at each $X$-position, see \autoref{fig:VelocityProfile}(a).
For all mixtures, characterized by $\alpha$, the function $V_\mathrm{avg}(X)$ displays a maximum followed by a decline (\autoref{fig:VelocityProfile}(b)) which is similar to granular flow down an inclined plane. An increase or decrease of $V_\mathrm{avg}(X)$ corresponds to an acceleration or deceleration of the particles in the flow. While spherical particles ($\alpha=0$) reveal 
a relatively sharp velocity profile, the addition of even a small proportion of non-spherical particles changes the flow significantly. The velocity drop reduces as the proportion of non-spherical particles, $\alpha$, increases. Thus, an increase in surface flow velocity is observed with increasing $\alpha$. This occurs due to the flow of particles down a stiffer slope of the heap in front of the moving blade as $\alpha$ increases.

\subsection{Spatial evolution of powder bed temperature}

\begin{figure}[htb!]
\centering
\begin{subfigure}{0.5\linewidth}
    \includegraphics[trim={0cm 0cm 0cm 0cm},clip,width=\linewidth]{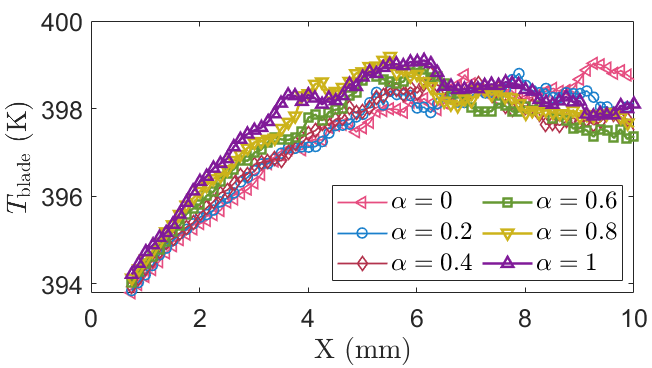}
\end{subfigure}
\caption{Temperature of the powder layer near the moving blade as a function of the spreading length, $X$, for mixtures of different composition $\alpha$.}
\label{fig:VelocityTemperatureSpace} 
\end{figure}

\autoref{fig:VelocityTemperatureSpace} 
shows the average particle temperature, $T_\text{blade}$, near the moving blade for mixtures of different compositions, characterized by $\alpha$. The slope $\text{d}T_\text{blade}/\text{d}X$ increases with $\alpha$ since the heat transfer rate is higher for granulate with a larger fraction of non-spherical particles due to their increased contact area.

While the temperature of a granulate of spheres ($\alpha=0$) increases continuously until eventually reaching its steady-state value, the introduction of even a small fraction of non-spherical particles ($\alpha>0$) leads to a notable change: Here, we observe a maximum before the temperature starts decreasing. The position of the maximum shifts to smaller $X$ with increasing $\alpha$. This can be understood from the more intense surface flow of spherical particles compared to non-spherical particles. Intense surface flow transports larger amounts of cold particles from the top of the heap to settle farther ahead of the blade and, thus, $X$, leading to the characteristic profiles shown in \autoref{fig:VelocityTemperatureSpace}.

\subsection{Homogeneity of the deposited powder layer}
Granular mixtures of particles of different shape tend to segregation. While this is a general phenomenon observed in granulate under various types of external excitation \cite{ketterhagen2008modeling,alizadeh2017effect}, it has been also reported in granulate subjected to shear \cite{cunez2024particle,kumar2024shape}. In powder spreading applications for additive manufacturing, segregation is undesired since it leads to inhomogeneous deposited layers and, thus, inhomogeneous macroscopic properties. Therefore, considerable efforts are made to achieve homogeneous powder layers. In this section, we study the homogeneity of the deposited layer of particles with respect to the particles' shape.

\subsubsection{Homogeneity along the spreading direction}

\begin{figure}[htb!]
\centering
\begin{subfigure}{0.48\linewidth}
    \includegraphics[trim={0cm 0cm 0cm 0cm},clip,width=\linewidth]{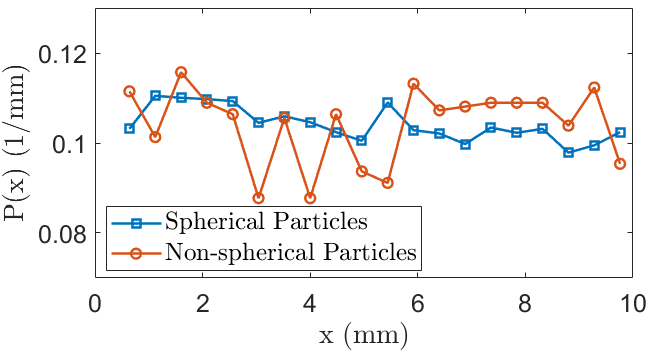}
    \subcaption{$\alpha = 0.2$}
\end{subfigure}
\hfill
\begin{subfigure}{0.48\linewidth}
    \includegraphics[trim={0cm 0cm 0cm 0cm},clip,width=\linewidth]{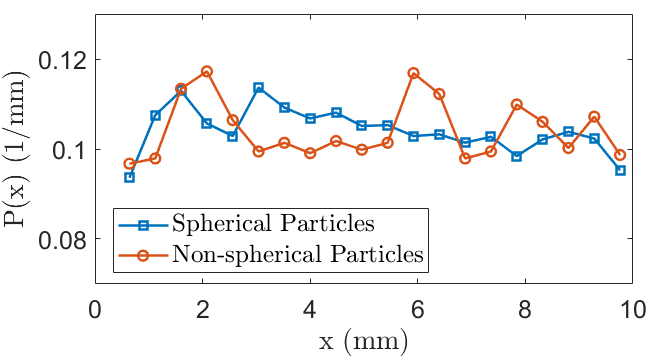}
    \subcaption{$\alpha = 0.4$}
\end{subfigure}
\hfill
\begin{subfigure}{0.48\linewidth}
    \includegraphics[trim={0cm 0cm 0cm 0cm},clip,width=\linewidth]{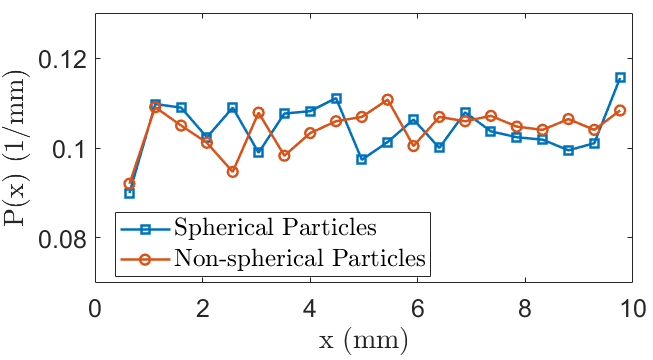}
    \subcaption{$\alpha = 0.6$}
\end{subfigure}
\hfill
\begin{subfigure}{0.48\linewidth}
    \includegraphics[trim={0cm 0cm 0cm 0cm},clip,width=\linewidth]{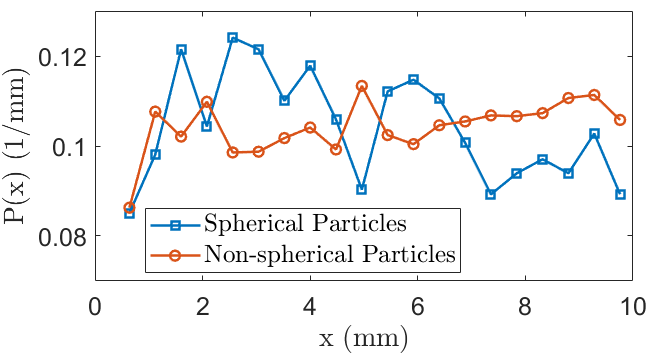}
    \subcaption{$\alpha = 0.8$}
\end{subfigure}
\caption{Number density of particles $P(x)$ along the spreading direction, for both non-spherical and spherical particles. The density was normalized independently for both types of particles.}
\label{fig:SegregationX}
\end{figure}

To characterize the homogeneity of the layer, \autoref{fig:SegregationX} 
shows the number density of particles along the spreading direction, $P(x)$, for some compositions characterized by $\alpha\in\{0.2,\dots,0.8\}$. Since the particle diameters are bound by a narrow interval, $[29.2,30]\,\mu\text{m}$, the particle number distribution is approximately proportional to the mass distribution. $P(x)$ is normalized such that $\int_0^\infty P(x)\,\text{d}x = 1$, similar to a probability density function. The normalization was done independently for both types of particles.

For all values of $\alpha$ we obtain relatively homogeneous distribution of both types of particles along the spreading direction, $P(x)\approx [0.1\pm0.015]\text{mm}^{-1}$. For $\alpha=0.2$, the fluctuations of the non-spherical particles is large due to the small total amount of non-spherical particles in the mixture. Similarly, for $\alpha=0.8$, the fluctuations of the spherical particles are large.

\subsubsection{Particle distribution across the depth of powder layer}

\begin{figure}[htb!]
\centering
\begin{subfigure}{0.49\linewidth}
    \includegraphics[trim={0cm 0cm 0cm 0cm},clip,width=\linewidth]{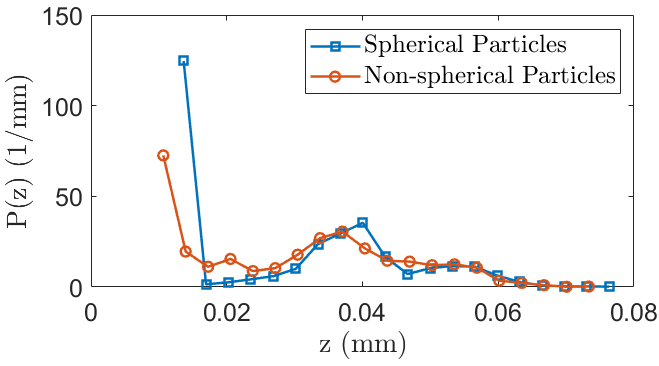}
    \subcaption{$\alpha = 0.2$}
\end{subfigure}
\hfill
\begin{subfigure}{0.49\linewidth}
    \includegraphics[trim={0cm 0cm 0cm 0cm},clip,width=\linewidth]{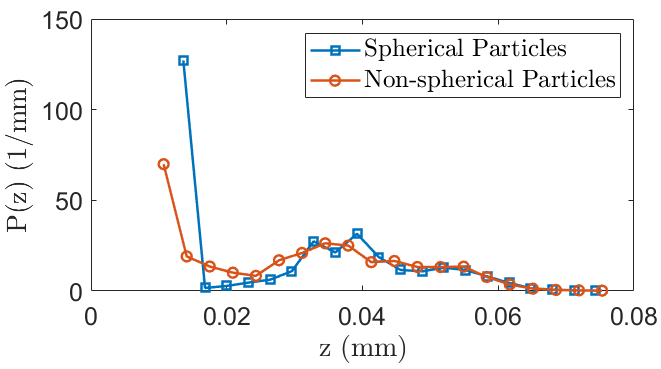}
    \subcaption{$\alpha = 0.4$}
\end{subfigure}
\hfill
\begin{subfigure}{0.49\linewidth}
    \includegraphics[trim={0cm 0cm 0cm 0cm},clip,width=\linewidth]{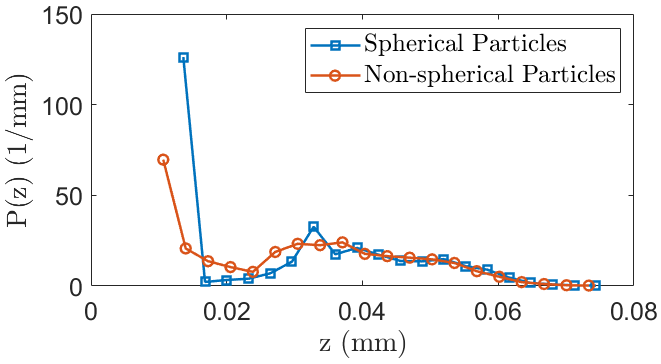}
    \subcaption{$\alpha = 0.6$}
\end{subfigure}
\hfill
\begin{subfigure}{0.49\linewidth}
    \includegraphics[trim={0cm 0cm 0cm 0cm},clip,width=\linewidth]{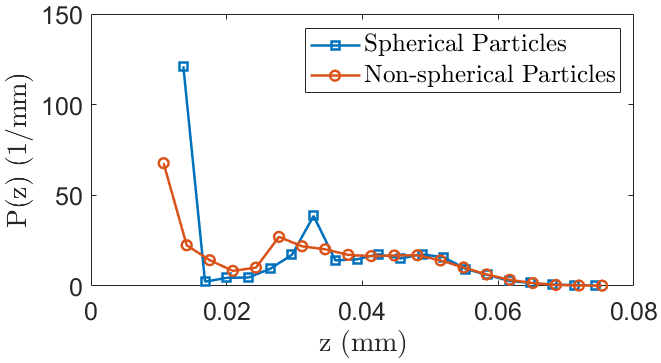}
    \subcaption{$\alpha = 0.8$}
\end{subfigure}
\caption{Particle number distribution with respect to the height $z$ of the powder for powders of various composition, $\alpha$. The distribution of spherical and nonspherical particles have ben normalized separately}
\label{fig:SegregationZ}
\end{figure}

The particle number density with respect to the height $z$ of the powder layer is shown in \autoref{fig:SegregationZ}. While the distance of the center of mass of the spheres always peaks at $0.015$ mm, equivalent to their radius, the distance of the center of mass of non-spherical particles is smaller because these particles orient themselves to minimize their potential energy. This effect is evident in \autoref{fig:SegregationZ}(a-d).

\section{Summary}
We studied the spread of binary mixtures of spherical and nonspherical particles of PA12 powder in a process of additive manufacturing by means of DEM where we took into account (a) the dependence of the material characteristics on temperature, and (b) the heat transfer due to thermoconductivity and radiation. Taking temperature and heat transfer into account is important for additive manufacturing processes, as the temperature changes considerably during the coating process and can be close to the melting temperature.

Changing the composition of the mixtures, characterized by the volume fraction of nonspherical particles, $\alpha$, influences the shape of the material (the heap in front of the blade) and the velocity profile near the surface while it has much less effect on the distribution of particles in the deposited powder layer.




\section*{Acknowledgement}
We acknowledge Holger G\"otz for discussions. This work was supported by the Interdisciplinary Center for Nanostructured Films (IZNF), the Competence Unit for Scientific Computing (CSC), and the Interdisciplinary Center for Functional Particle Systems (FPS) at Friedrich-Alexander-Universität Erlangen-Nürnberg.

\section*{Conflicts of interests}
There are no conflicts to declare.

\bibliography{references}

\end{document}